\pdfoutput=1
\documentclass{article}
\usepackage{locata,amsmath,graphicx,url,times,amsfonts}
\usepackage{math}
\usepackage{enumitem}
\usepackage{algorithm}
\usepackage{algorithmic}
\usepackage{subfig}
\usepackage{amssymb}
\usepackage{multirow}
\usepackage{textcomp}

\newcommand{\inv}{^{\raisebox{.2ex}{$\scriptscriptstyle-1$}}}
\title{A Cascaded Multiple-Speaker Localization and Tracking System}

   \name{Xiaofei Li,$^{1}$
      Yutong Ban,$^{1}$
      Laurent Girin,$^{1, 2}$
      Xavier Alameda-Pineda,$^{1}$ 
      Radu Horaud,$^{1}$\thanks{This work was supported by the ERC Advanced Grant VHIA \#340113.}
      }
\address{$^1$ Inria Grenoble Rh\^one-Alpes, France\\ 
         $^2$ Univ. Grenoble Alpes, Grenoble-INP, GIPSA-lab, France\\
}
\begin{document}

\ninept
\maketitle

\begin{sloppy}

\begin{abstract}
This paper presents an online multiple-speaker localization and tracking method, as the INRIA-Perception contribution to the LOCATA Challenge 2018. First, the recursive least-square method is used to adaptively estimate the direct-path relative transfer function as an interchannel localization feature. The feature is assumed to associate with a single speaker at each time-frequency bin. Second, a complex Gaussian mixture model (CGMM) is used as a generative model of the features. The weight of each CGMM component represents the probability that this component corresponds to an active speaker, and is adaptively estimated with an online optimization algorithm. Finally, taking the CGMM component weights as observations, a Bayesian multiple-speaker tracking method based on the variational expectation maximization algorithm is used. The tracker accounts for the variation of active speakers and the localization miss measurements, by introducing speaker birth and sleeping processes. The experiments carried out on the development dataset of the challenge are reported. 
\end{abstract}

\begin{keywords}
sound-source localization, multiple moving speakers, tracking, reverberant environments, LOCATA Challenge
\end{keywords}

\section{Introduction}
\label{sec:intro}

For multiple-speaker localization, the W-disjoint orthogonality (WDO) \cite{yilmaz2004} assumption is widely used. It assumes that the audio signal is dominated by only one speaker in each small region of the time-frequency (TF) domain, because of the natural sparsity of speech signals in this domain. After applying the short-time Fourier transform (STFT), interchannel localization features, e.g. interaural phase difference, can be extracted. To assign the interchannel features to multiple speakers, a mixture of Gaussian mixture models (GMMs) is used as a generative model of the interchannel features of multiple speakers in \cite{mandel2010} with each GMM representing one speaker, and then feature assignment was done based on the maximum likelihood criteria. 
In \cite{dorfan2015}, instead of setting one GMM for each speaker, a single complex GMM (CGMM) is used for all speakers with each component representing one candidate speaker location. After maximizing the likelihood of the  features, with an expectation-maximization (EM) algorithm, the weight of each component represents the probability that there is an active speaker at the corresponding candidate location. To localize moving speakers, in \cite{schwartz2014}, based on a CGMM model similar to \cite{dorfan2015}, a recursive EM algorithm was proposed to update online the CGMM component weights. Counting and localization of active speakers can be jointly carried out by selecting the components with large weights.
Taking the instantaneous outputs of a localization method as observations and using a speaker dynamic model, Bayesian tracking techniques estimate the posterior distribution of source locations, e.g. \cite{vermaak2001nonlinear,valin2007robust}.
To tackle the tracking problem with an unknown and time-varying number of speakers, additional model features such as observation-to-speaker assignments, speaker track birth/death processes and a model of speech activity can be included \cite{evers2015,ba2016line,gebru2017audio,ban2017}. 

In the present paper, we present our online multiple-speaker localization and tracking method contributed to the LOCATA Challenge, which is composed of three modules: 
\begin{itemize}[leftmargin=*]
 \item A recursive DP-RTF (direct-path relative transfer function) estimation module. In the STFT domain, the room impulse response (RIR) can be approximated by the convolutive transfer function (CTF) \cite{avargel2007,talmon2009}. DP-RTF is defined by the ratio between the first taps of the CTF of two microphones, thus encodes the direct-path information and is used as an interchannel feature being robust against reverberation. The CTF estimation used for DP-RTF extraction was formulated in batch mode in \cite{li2016taslp,li2017taslp} with speakers considered as static. Based on recursive least-square (RLS), the online CTF estimation method was proposed in \cite{li2018sam,li:hal-01851985}, and will be briefly presented in this paper, for moving speaker localization. 
 \item An online multiple-speaker localization module. In \cite{li2017taslp}, we adopt the above-mentioned CGMM model \cite{dorfan2015} to assign the DP-RTF features to speakers, in addition, an entropy-based regularization term was used to impose the spatial sparsity of the estimated component weights. However, \cite{li2017taslp} only considered the batch mode and static speakers. The recursive EM algorithm proposed in \cite{schwartz2014} was adopted in \cite{li2018sam} for online liklihood maximization without using the entropy regularization. Furtherly, an online optimization method, i.e. exponentiated gradient (EG) \cite{kivinen1997}, was used in \cite{li:hal-01851985} for simultaneously online liklihood maximization and entropy minimization. This method will be briefly presented in this paper.
 \item A multiple-speaker tracking module. In \cite{li:hal-01851985}, a  multiple-speaker tracking method was proposed. The results of the above localization module are taken as inputs and efficiently exploited in a Bayesian framework. The problem is efficiently solved by a variational expectation maximization (VEM) algorithm. Speaker birth and sleeping processes are included in the tracking process. The sleeping process is efficient to tackle a missed detection by the localization procedure. This paper will briefly present this multiple-speaker tracker,  please refer to \cite{ba2016line,ban2017,ban2017tracking,li:hal-01851985} for more detailed algorithmic derivation and description.
\end{itemize}
In Fig. \ref{fig:flowchart}, the three layers of temporal evolutions depicts these three modules, respectively. In the following, we present them one by one from Section \ref{sec:dprtf} to \ref{sec:tracking}, and then gives the experiments on the development dataset of the LOCATA challenge \cite{LOCATA2018a} in Section \ref{sec:experiments}.

\begin{figure}[t]
\centering
{\includegraphics[width=0.41\textwidth]{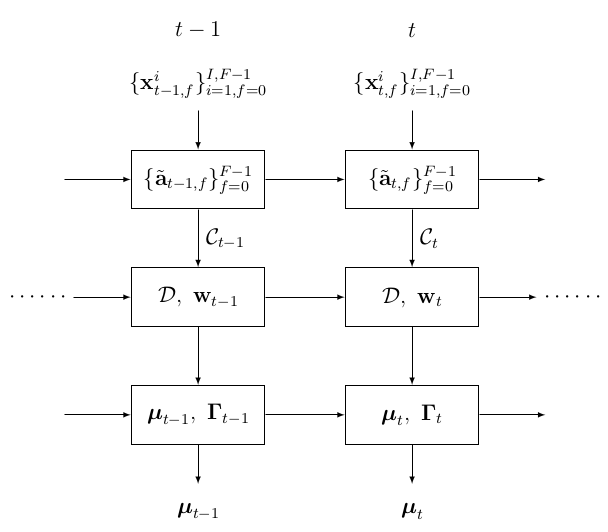}} 
\caption{\small{Flowchart of the three localization and tracking modules.}  } 
\label{fig:flowchart}
\vspace{-0.0cm}
\end{figure}

\section{Recursive DP-RTF estimation}\label{sec:dprtf}

To simplify the presentation, let us first consider the noise-free single-speaker case. In the time domain, the $i$-th, $i=1,\dots,I$, microphone signal is: $x^i(n)= a^i(n) \star s(n)$, where $n$ is the time index, $s(n)$ is the source signal, $a^{i}(n)$ is the RIR from the source to the $i$-th microphone, and $\star$ denotes the convolution. Applying the STFT, and using the CTF approximation, we have for each frequency index $f=0,\dots,F-1$: 
 $x^i_{t,f}=a^i_{t,f} \star s_{t,f}$, where $x^i_{t,f}$ and $s_{t,f}$ are the STFT coefficients of the corresponding signals, and the CTF $a^i_{t,f}$ is a subband representation of $a^{i}(n)$. Here, the convolution is executed w.r.t the frame index $t$. The first CTF coefficient $a^i_{0,f}$ mainly consists of the direct-path information, and DP-RTF is defined as the ratio between the first CTF coefficients of two channels:  ${a^i_{0,f}}/{a^r_{0,f}}$, where channel $r$ is the reference channel. 

Based on the cross-relation method \cite{xu1995}, for one microphone pair $(i,j)$,  we have $\mathbf{x}_{t,f}^{i \top}\mathbf{a}^j_f=\mathbf{x}_{t,f}^{j \ \top}\mathbf{a}^i_f$,  
where $\tp$ denotes matrix/vector transpose, the convolution vectors are $\mathbf{a}^i_f=[a^i_{0,f},\dots,a^i_{Q-1,f}]\tp$ and $\mathbf{x}^i_{t,f}=[x^i_{t,f},\dots,x^i_{t-Q+1,f}]\tp$, with $Q$ denoting the CTF length. We concatenate the CTF vector of all channels as $\mathbf{a}_f=[\mathbf{a}_f^{1 \top},\dots,\mathbf{a}_f^{I \top}]\tp$.
For each microphone pair $(i,j)$, we construct a cross-relation equation in terms of $\mathbf{a}_f$. For this aim, define: 
 \begin{align}\label{eq:xij}
\mathbf{x}^{ij}_{t,f} = [\underbrace{{0},\dots,{0}}_{(i-1)Q},  \mathbf{x}_{t,f}^{j \ \top}, \underbrace{{0},\dots,{0}}_{(j-i-1)Q}, -\mathbf{x}_{t,f}^{i \ \top}, \underbrace{{0},\dots,{0}}_{(I-j)Q}]\tp. 
 \end{align}  
 Then we have $\mathbf{x}_{t,f}^{ij \ \top} \mathbf{a}_f = 0$. The CTF vector $\mathbf{a}_f$ can be estimated by solving this equation. 
 To avoid a trivial solution, i.e. $\mathbf{a}_f=\mathbf{0}$, we constrain the first CTF coefficient of the reference channel, say $r=1$, to be equal to $1$. This leads to the following equation 
\begin{align}\label{eq:tildexija}
 \tilde{\mathbf{x}}_{t,f}^{ij \ \top} \tilde{\mathbf{a}}_f = y^{ij}_{t,f},
 \end{align}
 where 
$-y^{ij}_{t,f}$ is the first entry of $\mathbf{x}^{ij}_{t,f}$, 
 $\tilde{\mathbf{x}}^{ij}_{t,f}$ is $\mathbf{x}^{ij}_{t,f}$ with the first entry removed, and $ \tilde{\mathbf{a}}_f$ is the relative CTF vector:
 \begin{align}\label{eq:tak}
 \tilde{\mathbf{a}}_f = \left[\frac{\tilde{\mathbf{a}}_f^{1\top}}{a^{1}_{0,f}},\frac{\mathbf{a}_f^{2\top}}{a^{1}_{0,f}},\dots,\frac{\mathbf{a}_f^{I \top}}{a^{1}_{0,f}}\right]\tp.
 \end{align}
where  $\tilde{\mathbf{a}}^1_f=[a^1_{1,f},\dots,a^1_{Q-1,f}]\tp$, is $\mathbf{a}^1_f$ with the first entry removed.  For $i=2,\dots,I$, the DP-RTFs appear in (\ref{eq:tak}) as the first entries of $\frac{\mathbf{a}_f^{i \top}}{a^{1}_{0,f}}$. 
 
The DP-RTF estimation amounts to solving the linear problem Eq. (\ref{eq:tildexija}). Eq. (\ref{eq:tildexija}) is defined for one microphone pair at one frame. For the online case, we would like to update the estimate of $\tilde{\mathbf{a}}_f$ using the current frame, say $t$.  
There is a total of $M=I(I-1)/2$ distinct microphone pairs,
For notational convenience, instead of using $^{ij}$, we use $m=1,\dots,M$ denote the index of microphone pair. The fitting error of (\ref{eq:tildexija}) is $e^{m}_{t,f} = y^{m}_{t,f}-\tilde{\mathbf{x}}_{t,f}^{m \ \top} \tilde{\mathbf{a}}_{f}$. At the current frame $t$,  for the microphone pair $m$, RLS aims to minimize the error
\begin{align}\label{eq:cost}
 J^m_{t,f} = \sum_{t'=1}^{t-1}\sum_{m'=1}^M \lambda^{t-t'}|e^{m'}_{t',f}|^2+\sum_{m'=1}^m |e^{m'}_{t,f}|^2,
 \end{align}   
which sums up the fitting error of all the microphone pairs for the past frames and the microphone pairs up to $m$ for the current frame.
The forgetting factor $\lambda \in (0,1]$ gives exponentially lower weight to older frames, whereas at one given frame, all  microphone pairs have the same weight. In RLS, this minimization problem can be recursively solved. 
For the detailed recursion procedure please refer to Algorithm~1 in \cite{li:hal-01851985}. For each frame $t$, let $\tilde{\mathbf{a}}_{t,f}$ denote the CTF estimate, and $\tilde{c}^i_{t,f}, \ i=2,\dots,I$ denote the DP-RTF estimates extracted from $\tilde{\mathbf{a}}_{t,f}$. Note that implicitly we have $\tilde{c}^1_{t,f}=1$.

We now introduce how to extend the above method to the noisy multiple-speaker case. We assume that, over a short time, the speakers are static and only one source is active at each frequency bin. Therefore, the CTF can be estimated using a small number of recent frames. This can be done by adjusting the forgetting factor $\lambda$: To approximately have a memory of $P$ frames, we can set $\lambda=\frac{P-1}{P+1}$. $P$ should be empirically set to achieve a good tradeoff between the validity of the above assumptions and a robust CTF estimate. 
To suppress the noise, we use the inter-frame spectral subtraction algorithm proposed in \cite{li2015icassp,li2018sam}. At each frequency bin $f$, frames are first classified into speech frames and noise frames, then inter-frame spectral subtraction is applied between them. In the RLS process, only the speech frames (after spectral subtraction) are used, and the noise frames are skipped. In practice, a DP-RTF estimate can sometimes be unreliable due to the imperfect performance of the above-mentioned methods. We use the consistency test method proposed in \cite{li2017taslp,li:hal-01851985} to detect the unreliable estimates.  
Finally, at frame $t$, we obtain a set of features  $\mathcal{C}_t=\{\{\hat{c}^i_{t,f}\}_{i\in\mathcal{I}_f}\}_{f=0}^{F-1}$, where $\mathcal{I}_f\subseteq\{2,\dots,I\}$ denotes the set of microphone indices that pass the consistency test. Each of the features is assumed to be associated with a single speaker.
 
\section{Online Multiple-Speaker Localization} \label{sec:eg}

In order to assign the DP-RTF features in $\mathcal{C}_t$ to speakers, the CGMM generative model proposed in \cite{dorfan2015} is adopted.
We define a set $\mathcal{D}$ of $D$ candidate source locations. Let $d=1,\dots,D$ denote the location index. 
The probability, that an observed feature $\hat{c}^i_{t,f}$ is emitted
by candidate locations, is modelled by a CGMM: 
\begin{equation}\label{eq:CGMM}
P (\hat{c}^i_{t,f}| \mathcal{D}) = \sum_{d=1}^Dw^d\mathcal{N}_c(\hat{c}^i_{t,f};c_f^{i,d},\sigma^2),
\end{equation}
where the mean $c_f^{i,d}$ is precomputed based on the direct-path propagation model from the $d$-th candidate location to the microphone $i$. The variance $\sigma^2$ is empirically set as a constant value.
The component weight $w^d \geq 0$, with $\sum_{d=1}^Dw^d=1$, is the only free model parameter. Let us denote the vector of weights with $\wvect=[w^1, ..., w^D]\tp$. 

Let $\mathcal{L}_t$ denote the log-likelihood (normalized by the number of features) of the features in $\mathcal{C}_t$, as a function of $w^d$.
Once $\mathcal{L}_t$ is maximized, the component weight $w^d$ represents the probability that there exist an active speaker at the $d$-th candidate location. In addition, taking into account the fact that the number of actual active speakers is much lower than the number of candidate locations, an entropy minimization is taken as a regularization term to impose a sparse distribution of $w^d$. The entropy is defined as $ H = -\sum_{d=1}^Dw^d\text{log}(w^d)$. Then the overall cost function is $-\mathcal{L}_t+\gamma H$. In the present online framework, we proceed to a recursive update of the weight vector at current frame $t$, hence now denoted $\wvect_{t}$, based on the estimate at previous frame $\wvect_{t-1}$ and on the DP-RTF features of the current frame. This can be formulated as an online optimization problem \cite{kivinen1997}: 
\begin{align}\label{eq:oopt}
&{\wvect}_t = \mathop{\textrm{argmin}}_{\wvect} \ \chi(\wvect,\wvect_{t-1})+ \eta(-\mathcal{L}_t+\gamma H), 
\end{align}
with the constraints that $w^d, \ d =1,\dots,D$ are positive and sum to 1. $\chi(\wvect,\wvect_{t-1})$ is some distance measure between $\wvect$ and $\wvect_{t-1}$. The positive constant  $\eta$ controls the parameter update rate.  To exploit the fact that $w^d$ are probability masses, we use the Kullback-Leibler divergence, i.e. $\chi(\wvect,\wvect_{t-1})=\sum_{d=1}^D w^d\text{log}\frac{w^d}{w^d_{t-1}}$, which results in the exponentiated gradient algorithm \cite{kivinen1997}. Let $\Delta_{t-1}= \frac{\partial (\eta(-\mathcal{L}_t+\gamma H))}{\partial w^d}\Big|_{w^d_{t-1}}$ denote the partial derivatives of $\eta(-\mathcal{L}_t+\gamma H)$ w.r.t $w^d$ at the point $w^d_{t-1}$. 
Then, the exponentiated gradient $r_{t-1}^d=e^{-\Delta_{t-1}}$
is used to update the weights  
\begin{align}\label{eq:eg}
 w^d_{t} = \frac{r_{t-1}^dw_{t-1}^d}{\sum_{d'=1}^D r_{t-1}^{d'}w_{t-1}^{d'}}, \qquad \ \forall d \in \mathcal{D}.
\end{align}
It is obvious from (\ref{eq:eg}) that the parameter constraints, namely positivity and summation to 1, are automatically satisfied.

\section{Multiple-Speaker Tracking}\label{sec:tracking}

In the following, upper case letters denote random variables while lower case letters denote their realizations. 
Let $N$ be the maximum number of tracks (speakers), and let $n$ be the speaker index. Moreover, let $n=0$ denote \textit{no speaker}, or background noise.  Let $\Svect_{tn}$ be a latent (or state) variable associated with speaker $n$ at frame $t$, and let $\Smat_t = (\Svect_{t1}, \dots, \Svect_{tn}, \dots, \Svect_{tN})$. $\Svect_{tn}$ is composed of two parts: the speaker direction and the speaker velocity. In this work, speaker direction is defined by an azimuth $\theta_{tn}$. To avoid phase (circular) ambiguity we describe the direction with the unit vector 
$\Uvect_{tn} = (\cos(\theta_{tn}), \sin(\theta_{tn}))^{\top}$. 
Moreover, let  $V_{tn}\in\mathbb{R}$ be the angular velocity. 
Altogether we define a realization of the state variable as $\svect_{tn} = [\uvect_{tn} ; v_{tn}]$ where the notation $[\cdot ; \cdot]$ stands for vertical vector concatenation.

As mentioned above, the CGMM component weight $w^d$ represents the probability that there exist an active speaker at the $d$-th candidate location. The frame-wise localization of active speakers can be carried out by peak picking over $w^d$. However, to fully use the weight information, without applying peak picking, all the candidate locations and their associated weights are used. Formally, 
let $\Omat_t = (\Ovect_{t1}, \dots, \Ovect_{td}, \dots, \Ovect_{tD})$ be the observed variables at frame $t$. Each realization $\ovect_{td}$ of $\Ovect_{td}$ is composed of a candidate location, or azimuth $\tilde{\theta}_{td}\in\mathcal{D}$, and a weight $w_{td}$. As above, let the azimuth be described by a unit vector $\bvect_{td}=(\cos(\tilde{\theta}_{td}), \sin(\tilde{\theta}_{td}))^{\top}$. In summary we have $\ovect_{td} = [\bvect_{td} ; w_{td}]$.
Moreover, let  $Z_{td}$ be a (latent) assignment variable associated with each observed variable $\Ovect_{td}$, such that $Z_{td}= n$ means that the observation indexed by $d$ at frame $t$ is assigned to speaker $n\in\{0,\dots, N\}$.

\subsection{Bayesian Tracking Model}

The problem at hand can be cast into the estimation of the filtering distribution $p(\smat_t, \zvect_t | \omat_{1:t})$, and further inference of $\smat_t$ and $\zvect_t$. 
By applying the Bayes rule, the filtering distribution is proportional to:
\begin{align}
p(\smat_t, \zvect_t | \omat_{1:t}) \propto  p(\omat_t|  \smat_t, \zvect_t) p(\zvect_t) p(\smat_t | \omat_{1:t-1}),
\label{eq:posterior_Bayes}
\end{align}
which contains the following three terms.

The audio observation model $p(\omat_t|  \smat_t, \zvect_t)$ describes the distribution of the observations given speakers state and assignment. 
We assume the different observations are independent conditionally to speakers state and assignment.
For each observation, we adopt the weighted-data GMM model of \cite{gebru2016algorithms}: 
$p (\bvect_{td} |  Z_{td}=n,  \smat_{tn}; w_{td})= 
\mathcal{N} (\bvect_{td};  \Mmat \svect_{tn} , \frac{1}{w_{td}}\Sigmamat)$ for $n \in \{1,\dots,N\}$, 
where the matrix $\Mmat =\left[ \Imat_{2 \times 2}, \textbf{0}_{2 \times 1} \right]$ projects the state variable onto the space of source directions and $\Sigmamat$ is a covariance matrix (set empirically to a fixed value). Note that the weight plays the role of a precision: The higher the weight $w_{td}$, the more reliable the source direction $\bvect_{td}$.  
The case $Z_{td} =0$ follows a uniform distribution over the volume of the observation space, i.e. $p (\bvect_{td} |  Z_{td}=0)= 
\mathcal{U} (\textrm{vol} (\mathcal{G}))$.

The prior distribution of the assignment variable, i.e. $p(\zvect_t)$, is independent over observations and is assumed to be uniformly distributed over all the speakers, i.e. $p(Z_{td}=n)=\pi_{dn}= \frac{1}{N+1}$. 

To calculate the state predictive distribution $p(\smat_t | \omat_{1:t-1})$, we marginalize it over $\smat_{t-1}$:
\begin{align}
\label{eq:predictive distribution}
p(\smat_t | \omat_{1:t-1}) &= \int p(\smat_t | \smat_{t-1}) p( \smat_{t-1} | \omat_{1:t-1}) d\smat_{t-1}.
\end{align}
We model the state dynamics $p(\smat_t | \smat_{t-1})$ as a linear-Gaussian first-order Markov process, independent over the speakers, i.e. $p(\smat_{t,n} | \smat_{t-1,n}) = \mathcal{N}
(\svect_{tn};\Dmat_{t-1,n} \svect_{t-1,n},\Lambdamat_{tn})$,  
where $\Lambda_{tn}$ is the dynamics' covariance matrix and $\Dmat_{t-1,n}$ is the state transition matrix. Importantly, since the state position subvector $\uvect_{tn}$ lies on the unit circle, the dynamic model is designed for circular motion. Given the estimated azimuth angle $\theta_{t-1,n}$ at frame $t-1$,  the state transition matrix can be written as:
\begin{equation}
\label{eq:transition-matrix}
\Dmat_{t-1,n} = \begin{pmatrix}
1 & 0 &-\sin(\theta_{t-1,n}) \\
0 & 1 & \cos(\theta_{t-1,n}) \\
0 & 0 & 1 \\
\end{pmatrix}.
\end{equation}
In the following $\Dmat_{t-1,n}$ is written as $\Dmat$ for notational simplicity. 

\subsection{Variational Expectation Maximization Algorithm}
To estimate the model parameters $\Thetavect$, and infer the posteriori distribution of ($\smat_t, \zvect_t$), we adpot an EM algorithm that alternates between computing and maximizing the expected complete-data log-likelihood:
\begin{equation}
J(\Thetavect,\Thetavect^{o}) = \mathbf{E}_{p(\zvect_t,\smat_t|\omat_{1:t},\Thetavect^{o})} \left[\log p(\zvect_t,\smat_t,\omat_{1:t}|\Thetavect) \right],
\label{eq:q-func}
\end{equation}
where $\mathbf{E}$ denotes expectation, $\Thetavect^{o}$ are the ``old'' model parameter estimates (obtained at previous iteration). Given the hybrid combinatorial-continuous nature of the latent space, it is computationally heavy to operate with the exact a posterior distribution $p(\zvect_t,\smat_t|\omat_{1:t})$.
We thus use a variational approximation to solve the problem \cite{ba2016line}, which factorizes the posterior distribution as:
\begin{equation}
\label{eq:variational-approximation}
p(\zvect_t,\smat_t|\omat_{1:t}) \approx 
q(\zvect_t,\smat_t) = q(\zvect_{t}) 
\prod_{n=0}^{N} q(\svect_{tn}).
\end{equation} 
It is seen that the posterior distribution factorizes across speakers. This principle is also valid for time $t-1$. As a result, the predictive distribution (\ref{eq:predictive distribution}) also factorizes across speakers. If the posterior distribution $q(\svect_{t-1,n})$ is assumed to be a Gaussian with mean $\muvect_{t-1,n}$ and variance $\Gammamat_{t-1,n}$, 
then the per-speaker predictive distribution $p(\svect_{tn}|\omat_{1:t-1})$ is a Gaussian:  
\begin{align}\label{eqn:predictive-motion-mod}
p(\smat_{tn}|\omat_{1:t-1}) =  \mathcal{N}(\smat_{tn};\Dmat\muvect_{t-1,n},\Dmat\Gammamat_{t-1,n}\Dmat^{\top}+\Lambdamat_{tn}).
\end{align}

In E-S step, it can be derived that the variational posterior distribution of $q(\svect_{tn})$ is a Gaussian distribution, with variance and mean respectively as 
\scriptsize
\begin{align}
\Gammamat_{tn} & =  \ \Big[\Big(\sum_{d=1}^{D}\alpha_{tdn} w_{t}^{d}\Big) \Hmat^{\top}{\Sigmamat}\inv\Hmat 
 + \Big(\Lambdamat_{tn}+\Dmat\Gammamat_{t-1,n}\Dmat^{\top}\Big)\inv \Big]\inv,  \nonumber \\
\muvect_{tn}  & =  \Gammamat_{tn} \Big[ \Hmat^{\top}{\Sigmamat}\inv \Big(\sum_{d=1}^{D}\alpha_{tdn} w_{t}^{d} \bvect_{td}\Big)   
    + \Big(\Lambdamat_{tn}+\Dmat\Gammamat_{t-1,n}\Dmat^{\top}\Big)\inv\Dmat\muvect_{t-1,n}\Big], \nonumber
\end{align}
\small
where $\alpha_{tdn}= q(Z_{td} = n)$ is the variational posterior distribution of the assignment variable, which is derived in E-Z step, the model parameters $\Lambdamat_{tn}$ is udated in M-step, please refer to \cite{li:hal-01851985} for detailed descriptions.
At time $t$, VEM converges after a few iterations, e.g. 5 iterations used in this work. After convergency, the posterior mean of $\svect_{tn}$, i.e. $\muvect_{t}=(\muvect_{t1},\dots,\muvect_{tn},\dots,\muvect_{tN})$, is output as the result of speaker tracking, and together with the posterior variance $\Gammamat_{t}=(\Gammamat_{t1},\dots,\Gammamat_{tn},\dots,\Gammamat_{tN})$ are transmitted to time $t+1$.

\subsection{Speaker Track Birth Process and Activity Detection}
A track birth process is used to initialize new tracks, i.e. new speakers that enter the scenario. The general principle is the following. In a short period of time, say from frame $t-L$ to frame $t$, we assume that at most one new (yet untracked) speaker appears. For each frame from $t-L$ to $t$, we select the observation assigned to the background noise with the highest weight, and obtain an observation sequence $\tilde{\ovect}_{t-L:t}$. We then calculate the marginal likelihood of this sequence according to our model, $\tau_0=p(\tilde{\ovect}_{t-L:t})$.  If these observations have been generated by a new speaker, they exhibit smooth trajectories, and $\tau_0$ will be high. Therefore, the birth process is conducted by comparing $\tau_0$ with a threshold. The posterior distribution of the assignment variable, i.e. $\alpha_{idn}$, can be used for multiple-speaker activity detection. This can be formalized as testing, for each frame $t$ and each speaker $n$, whether the weighted assignments$\sum_{d = 1}^{D} \alpha_{idn} w_{i}^{d}$ (averaged over a small number of frames) is larger than a threshold. 

\section{Experiments on LOCATA Development Data}\label{sec:experiments}

\begin{figure*}[t]
\centering
\subfloat[Ground truth]{\includegraphics[width=0.32\textwidth]{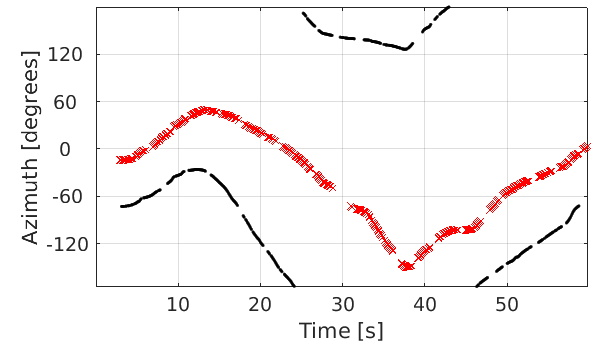}} 
\subfloat[Localization: GMM Weights]{\includegraphics[width=0.32\textwidth]{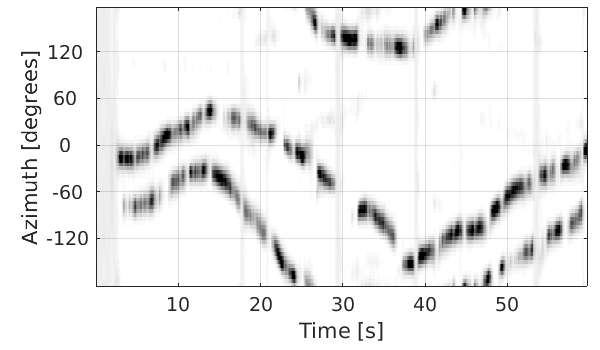}} 
\subfloat[Tracking]{\includegraphics[width=0.32\textwidth]{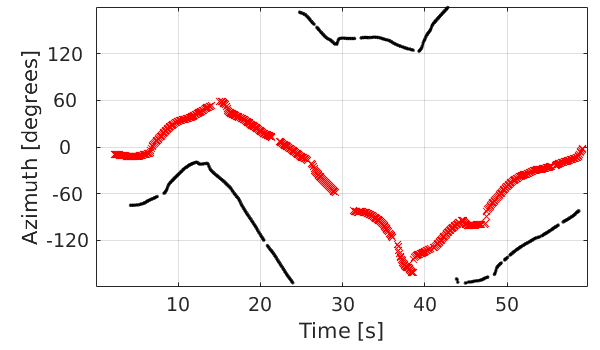}} \\
\caption{\small{Results of speaker localization and tracking for Recording \#1 / Task \#6. (a) Ground truth trajectory and voice activity (red for speaker 1, black for speaker 2). Intervals in the trajectories are speaking pauses. (b) Result for localization, i.e. GMM Weights. (f) Result for   tracker. Black and red colors demonstrate a succesful tracking, i.e. continuity of the tracks despite of speech pauses.}}
\label{fig:locata_example}
\vspace{-0.0cm}
\end{figure*}    

We report the results of on the LOCATA development corpus for tasks \#3 and \#5 with a single moving speaker, and tasks \#4 and \#6 with two moving speakers, each task comprising three recorded sequences. In this work, we use four microphones with indices $\{5,8,11,12\}$
out of the twelve microphones of a spherical array built in the head of a humanoid robot, i.e. NAO, to perform azimuth localization and tracking. These four microphones are mounted on the top of the robot head, and they approximately lie in a horizontal plane. 
We perform $360^\circ$-wide azimuth estimation and tracking: $D=72$ azimuth directions at every $5^\circ$ in [$-175^\circ$, $180^\circ$] are used as candidate directions. The TDOAs are computed based on the coordinate of microphones, which are then  used to compute the phase of the CGMM means, while the magnitude of the CGMM means are set to a constant  for all the frequencies. All the recorded signals are resampled to $16$ kHz. The STFT uses the Hamming window with length of $16$~ms and shift of $8$~ms. The CTF length is $Q=8$ frames. The RLS forgetting factor $\lambda$ is computed using $\rho=1$. The exponentiated gradient update factor is $\eta=0.07$. The entropy regularization factor is $\gamma=0.1$. 
For the tracker, the covariance matrix is set to be isotropic $\Sigmamat = 0.03\Imat_{2}$. \\
\textbf{A localization and tracking example} Fig.~\ref{fig:locata_example} shows an example for a LOCATA sequence. Two speakers are moving and continuously speaking with short pauses. Fig.~\ref{fig:locata_example}~(b) shows that the localization method achieves a heatmap with negligible  interferences  and smooth peak evolution. 
From Fig.~\ref{fig:locata_example}~(c), it is seen that the tracking method further smooth the speaker moving trajectories. Even when the observations have a low weight, the tracker is still able to give the correct speaker trajectories. This is ensured by exploiting the source dynamics model. 
As a result, the tracker is able to preserve the identity of speakers in spite of the (short) speech pauses. In the presented sequence example, the estimated speaker identities are quite consistent with the ground truth. \\
\textbf{Quantitative results} The following metrics are used for quantitative evaluation.
The detected speaker is considered to be successfully localized if the azimuth difference is not larger than $15^{\circ}$. The absolute error is calculated for the successfully localized sources. The mean absolute error (MAE) is computed by averaging the absolute error of all speakers and frames. For the unsuccessful localizations, we count the miss detections (MD) (speaker active but not detected) and false alarms (FA) (speaker detected but not active). Then the MD and FA rates are computed, using all the frames, as the percentage of the total MDs and FAs out of the total number of actual speakers, respectively. In addition to these localization metrics, we also count the identity switches (IDs) to evaluate the tracking continuity. IDs represents the number of the identity changes in the tracks for a whole test sequence. Table~\ref{tab:locata} gives the quantitative localization and tracking results. The localization results are obtained by applying peak picking on the GMM weights.      
The tracker slightly reduces the FA rate compared to the localization method alone mainly by eliminating some spurious peaks that are present in the localization outputs. It also reduces the MD rate since some correct speaker trajectories can be recovered even when the observations have (very) weak weights, as explained above. The identity switches are mainly due to the crossing of speaker trajectories, a hard case for the source dynamics model.

\begin{table}[t]
\centering
\caption{\small Localization and tracking results.}
\label{tab:locata}
\begin{tabular}{c | c  c   c c      }   \vspace{1mm}
	       & MD rate (\%)  & FA rate (\%)  & MAE ($^\circ$)   & IDs \\
Localization   &  24.1     & 12.7      & 4.0              & -   \\
Tracking   &  22.7     & 12.4 	   & 4.1              &  10 \\
\end{tabular}
\end{table} 

\section{Conclusion}\label{sec:conclusion}
In this paper, we presented the INRIA-Perception contribution to the LOCATA Challenge 2018. We combined i) a recursive DP-RTF feature estimation method, ii) an online multiple-speaker localization method, and iii) an multiple-speaker tracking method. The resulting framework provides online speaker counting, localization and consistent tracking (i.e. preserving speaker identity over a track in spite of intermittent speech production). 


\scriptsize
\bibliographystyle{IEEEtran}
\bibliography{citations}

\end{sloppy}
\end{document}